\documentclass{article}
\usepackage{arxiv}
\usepackage{cite}
\usepackage{amsmath,amssymb,amsfonts}
\usepackage{algorithmic}
\usepackage{graphicx}
\usepackage{textcomp}
\usepackage{xcolor}
\usepackage{subcaption}
\usepackage{url}
\usepackage[breaklinks=true]{hyperref}
\usepackage{color}
\usepackage{soul}
\def\BibTeX{{\rm B\kern-.05em{\sc i\kern-.025em b}\kern-.08em
T\kern-.1667em\lower.7ex\hbox{E}\kern-.125emX}}

\begin{document}

\title{Not Sure Your Car Withstands Cyberwarfare}

\author{
 Giampaolo Bella \\
  University of Catania\\
  Catania, Italy \\
  \texttt{giampaolo.bella@unict.it} \\
   \And
 Gianpietro Castiglione \\
  University of Catania\\
  Catania, Italy \\
  \texttt{gianpietro.castiglione@phd.unict.it} \\
  \And
 Sergio Esposito \\
  University of Catania\\
  Catania, Italy \\
  \texttt{sergio.esposito@unict.it} \\
  \And
 Mario Raciti \\
  IMT School for Advanced Studies Lucca\\
  Lucca, Italy \\
  \texttt{mario.raciti@imtlucca.it} \\
  \And
 Salvatore Riccobene \\
  University of Catania\\
  Catania, Italy \\
  \texttt{riccobene@unict.it} 
}

\date{}
\maketitle

\begin{abstract}
Data and derived information about target victims has always been key for successful attacks, both during historical wars and modern cyber wars. Ours turns out to be an era in which modern cars generate a plethora of data about their drivers, and such data could be extremely attractive for offenders. 
This paper seeks to assess how well modern cars protect their drivers' data. It pursues its goal at a requirement level by analysing the gaps of the privacy policies of chief automakers such as BMW and Mercedes with respect to the General Data Protection Regulation (GDPR). 
It is found that both brands are still imprecise about how they comply with a number of GDPR articles, hence compliance often results non-verifiable. 
Most importantly, while BMW exhibits slightly broader compliance, both brands still fail to comply with a number of relevant articles of the regulation. An interpretation of these findings is a non-negligible likelihood that your car may turn against you should cyberwarfare break out.
\end{abstract}


\definecolor{purple}{rgb}{1,0,1}
\newcommand{\kibitz}[2]{\ifnum\Comments=1\textcolor{#1}{#2}\fi}

\newcommand{\mr}[1]{\kibitz{blue}{[\textbf{Mario:} #1]}}
\newcommand{\gb}[1]{\kibitz{red}{[\textbf{Giamp:} #1]}}
\newcommand{\gc}[1]{\kibitz{orange}{[\textbf{Gianpietro:} #1]}}
\newcommand{\se}[1]{\kibitz{purple}{[\textbf{Sergio:} #1]}}
\newcommand{\sr}[1]{\kibitz{yellow}{[\textbf{Salvatore:} #1]}}
\newcount\Comments  
\Comments=1

\section{Introduction}
\label{sec:introduction}

History has demonstrated the importance of data and information to defeat an opponent at war on innumerable occasions, for example, to coordinate the operations among allies, as well as to locate and strike the targets at their weakest points. Considering this, the present paper observes that the digital versions of data and information would continue to be of utmost importance through cyberwarfare today --- ``it is going to be a colossal task for humanitarian law to provide any protection for data in an armed conflict leading to sporadic patterns of damage and casualties. 
When data is not duly protected during an armed conflict, it can cause grave paralysis to day-to-day civilian activities''~\cite{cirsd}.

It is widely known that modern vehicles deliver personalised services and are equipped with numerous connectivity features, hence continuously gathering extensive data at present, from location tracking to driver behaviour and in-car media usage.
Such ever-growing data proliferation, while enhancing the overall driving experience and vehicle functionalities, clearly raises data protection risks in general. 
Therefore, this paper seeks to assess the data protection measures currently applied
to the automotive sector. 
This goal can be pursued at various architectural levels, such as requirements, design and implementation, and this paper focuses on the level of requirements. In simpler terms, no technological car system can shield driver data properly if the very requirements that the system wants to meet are somewhat \textit{insufficient}.


It can be observed that the data protection requirements a modern car promises to comply with are normally written in the companion \textit{usage policy} or \textit{privacy policy}, which are textual documents, sometimes rather long, that cars recommend drivers to go and find on the Internet or display on large infotainment screens when these are available.
In a juridical landscape of data protection that is defined, perhaps globally, by the EU General Data Protection Regulation (GDPR) \cite{GDPR}, it can be deduced that the goal of this paper translates to assessing the privacy policies of modern car automakers in terms of their compliance with GDPR.

GDPR sets stringent standards for personal data protection, including transparency, user consent, and user control over their data \cite{mozillaItsOfficial}.
It is not obvious whether car manufacturers have fully aligned their data collection and processing practices with GDPR mandates.
Potential misalignments range from inadequate data handling disclosures to insufficient mechanisms for user consent and control, leading to potential privacy breaches. 
Instances of non-compliance do not seem to be merely isolated, and we have previously observed that they may represent a broader systemic issue within the automotive industry~\cite{autoin23}. 



A due remark is about the specific personal data that could be potentially exfiltrated from modern cars with offensive aims in the case of cyberwar. 
Example data, to say the least, may encompass the GPS coordinates of targets and, more in general, the car features that may suffer exploitable vulnerabilities, such as manipulations of indoor temperature to cause (serious) discomfort as well as alterations of engine RPMs, car speed and trajectories to endanger safety. Therefore, this paper targets the following research question:

\begin{quote}
RQ. \textit{How likely is it for your car to turn into a cyber weapon against yourself in case of cyberwarfare?} 
\end{quote}

We highlight that the methodology that will be followed below does not aim to be quantitative. 
Rather, it engages in deriving an answer to the research question, once again, in GDPR terms. 
If the driver were enabled to enjoy all GDPR articles, then we would answer negatively. Conversely, if the driver were found, for example, to be unable to exercise their right to erasure (of their personal data), we would then conclude with some likelihood that the car could be exploited as a cyber weapon.

To get a stab at answering its research question, this paper 
discusses a GDPR gap analysis of two of the top 10 car manufacturers in 2023 \cite{bestsellingcars2023}, namely BMW and Mercedes-Benz, to assess their compliance and identify possible common compliance shortcomings. It is found that the largest number of GDPR articles are not verifiable, a clear call for extra preciseness in the juridical language of the privacy policies of such leading representative automakers. 
Moreover, although it appears that BMW passes a larger number of controls, both brands still fail to pass a significant number of controls. 

\medskip
\textbf{\textit{Paper Outline.}} This paper follows a simple structure. Section~\ref{sec:related-work} outlines the related work; Section~\ref{sec:meth-gap} presents the methodology of the gap analysis applied between two car privacy policies and GDPR; Section~\ref{sec:disc} discusses the results; Section~\ref{sec:conc} concludes. 

\section{Related Work}
\label{sec:related-work}

This Section outlines the related work. The treatment is conveniently split into three subsections: Gap Analysis with respect to Automotive Standards focuses on studies in the compliance sphere of the automotive field; Fairness in Privacy Policies presents the concept of fairness across different facets; (Privacy) Risk Assessment for Smart Cars summarises attempts of modelling and analysing cybersecurity and privacy risks on board of connected and/or smart cars. As we shall detail below, to the best of our knowledge, there are no other works that conduct a gap analysis between the GDPR and privacy policies of top car manufacturers, and neither is there a work that focuses on privacy gap analysis results from a cyberwarfare perspective. These are the distinctive features of this paper.

\subsection{Gap Analysis with respective to Automotive Standards}

Siddiqui et al.~\cite{siddiqui} proposed a holistic cybersecurity engineering process that maps the requirements of ISO/SAE 21434~\cite{iso21434} onto traditional system design processes, encompassing phases from system generation to runtime. The proposed work systematically identifies, evaluates, protects, and manages cybersecurity risks throughout the automotive system life cycle, making it easier for automotive system designers to adhere to ISO/SAE 21434 standard standards.

Still, within the scope of ISO/SAE 21434, Kannan et al.~\cite{kannan} examined the gap between existing automotive Manufacturing Execution Systems (MES) and Industry 4.0 standards. The authors proposed a model-based requirements engineering approach to bridge this gap, analysing current MES tools and their compliance with standards like ANSI/ISA-95 and ANSI/ISA-88. The work identified discrepancies between standard requirements and the current software tools' compliance with those standards.

\subsection{Fairness in Privacy Policies}

Furthermore, Nagpal et al.~\cite{nagpal2018extracting} were among the first to conduct studies related to the fairness of a policy — the fairness level indicates how fair, proper and clean a text is, regarding the users' privacy concerns. The authors proposed a methodology to automatically extract a fairness value from public law documents leveraging semantic relatedness, namely the identification of some form of lexical or functional association between two words or concepts, based on the contextual or semantic similarity of those two words, regardless of their syntactical differences.

Bella et al.~\cite{aila} recalled the concept of fairness as the extent to which a policy statement respects natural persons’ privacy. The authors leveraged fairness to estimate the likelihood of threats deriving from privacy policies, in particular for the automotive domain. Yet, the work focused mainly on privacy policies, leaving additional standards and regulations like the GDPR out of scope.

\subsection{(Privacy) Risk Assessment for Smart Cars}

Wang et al.~\cite{Wang2021} proposed a threat-oriented risk assessment framework tailored for the smart car domain, with the aim, among others, of overcoming assumptions and subjectivity. This framework can be considered a precursor to ISO/IEC:21434~\cite{iso21434}, which was defined a year later. Also, the authors applied STRIDE and the attack tree method for the threat modelling. In addition, De Gusmão et al.~\cite{gusmao} proposed a risk analysis framework that adopts fault tree analysis. However, in both cases, the focus lies more on cybersecurity than privacy or GDPR-compliance aspects.

Moreover, Chah et al.~\cite{CHAH202236} applied the LINDDUN methodology~\cite{linddun} to elicit and analyse privacy requirements of CAV system, while respecting the privacy properties set by the GDPR. Such an attempt represents a solid baseline for the broader process of privacy risk assessment tailored for the smart car domain. Despite the application of LINDDUN, which tailored the analysis to privacy, the descriptions of the (limited and) available threats seem to predominantly focus on cybersecurity aspects rather than privacy and GDPR-compliance. In addition, the CAV system only represents a subset of the broader smart car domain.

Finally, Bella et al.~\cite{autoin23} advanced a framework for evaluating privacy risks in the automotive industry, specifically concerning data collected by cars. The authors use a double assessment approach, combining the asset-oriented ISO approach with the threat-oriented STRIDE~\cite{stride} approach. 
They analyse top-selling car brands plus Tesla, examining data categories collected, potential threats, and impacts of data breaches. The study aims to identify privacy risks and propose mitigation measures. However, it doesn't directly assess compliance with specific GDPR chapters or controls.

\section{Methodology of Gap Analysis}
\label{sec:meth-gap}

Arguably, modern cars' compliance with state-of-the-art privacy best practices needs to be improved.
As stated above, this paper aims to understand, in a fine-grained way, what requirements modern vehicles, and more in general, car manufacturers, mostly fail to comply with.
This Section details the process of conducting a gap analysis to assess the GDPR compliance of BMW and Mercedes-Benz based on their privacy policies. The analysis involved manual extraction and interpretation of GDPR requirements, followed by a compliance check against the brands’ public documentation.

\subsection{Extraction of GDPR Controls}

To assess GDPR compliance in the automotive domain, we manually extracted 91 controls on privacy requirements from the GDPR text, covering Articles 5 to 41. These articles were selected for two main reasons. First, for their relevance to the obligations of data controllers and processors to be considered fully compliant with the GDPR, and secondly for their applicability within the automotive domain. The controls are derived by transforming the GDPR requirements into interrogative form.
For example, Article 5 reads:

\begin{quote}
    \textit{``The data subject shall have the right to obtain from the controller confirmation as to whether or not personal data concerning him or her are being processed, and, where that is the case, access to the personal data and \dots''}
\end{quote}

Following our argument, Article 5 can be reformulated into interrogative form as:

\begin{quote}
    \textit{``Can the data subject exercise the right of access?''}.
\end{quote}

This transformation process allowed us to systematically identify compliance questions tailored to the automotive domain, ensuring a fine-grained evaluation. Also, this process led us to reconsider the resulting item, i.e., 
some paragraphs were condensed into smaller controls while others were extended, to appropriately highlight the privacy requirement that we wanted to extract.

\subsection{Selection of Representative Car Brands}

We selected BMW and Mercedes-Benz as representatives for this study based on their leading positions within the global automotive market in 2023~\cite{bestsellingcars2023} and their prominence in integrating connectivity features that naturally generate substantial amounts of data. Therefore, BMW and Mercedes-Benz provide a relevant setting for GDPR compliance analysis in the automotive sector.

\subsection{Collection of Car Brands' Privacy Policies}

We collected the privacy policies of the car brand by simply searching on the website of the brand, in particular, by searching the policies related to the car, since other privacy policies may exist, e.g., the ones related to the usage of the website itself.
We double-checked the documents that we selected with the privacy policy available on the smartphone app used to connect to the car' services.
In particular, for BMW we assessed the ``ConnectedDrive'' privacy policy \cite{BMWConnectDrive} and the legal notice. 
For Mercedes-Benz, we assessed the ``Data Protection Policy EU'' of the Mercedes-Benz Group \cite{MercedesPrivacyPolicy}.
While the document from Mercedes-Benz is a sort of company group policy rather than a real privacy policy meant for users, its contents are strongly oriented toward GDPR compliance. 

\subsection{Analysis of Privacy Policies Compliance with the GDPR}

We manually checked each control derived from the GDPR against the above-mentioned documents. The compliance of each control can be categorised into four groups:

\begin{itemize}
    \item \textbf{Passed}: the policy complies with the GDPR control.
    \item \textbf{Not Passed}: the policy does not satisfy the GDPR control.
    \item \textbf{Not Verifiable}: the policy lacks sufficient detail to verify compliance.
    \item \textbf{Not applicable}: the requirement is not applicable.
\end{itemize}

Table \ref{tab:compliance} summarises the raw results of the gap analysis performed on the car brands, i.e., BMW and Mercedes-Benz.

\begin{table}[ht]
\centering
\caption{Gap Analysis}
  \begin{tabular}{|l|c|c|}
    \hline
    & \textbf{BMW} & \textbf{Mercedes-Benz} \\
    \hline
    \textbf{Passed} & 29 & 21 \\ \hline
    \textbf{Not Passed} & 18 & 33 \\ \hline
    \textbf{Not Verifiable} & 39 & 37 \\ \hline
    \textbf{Not Applicable} & 5 & 0 \\ \hline
\end{tabular}
\label{tab:compliance}
\end{table}

\begin{figure}[ht]
    \centering
    \includegraphics[width=0.75\linewidth]{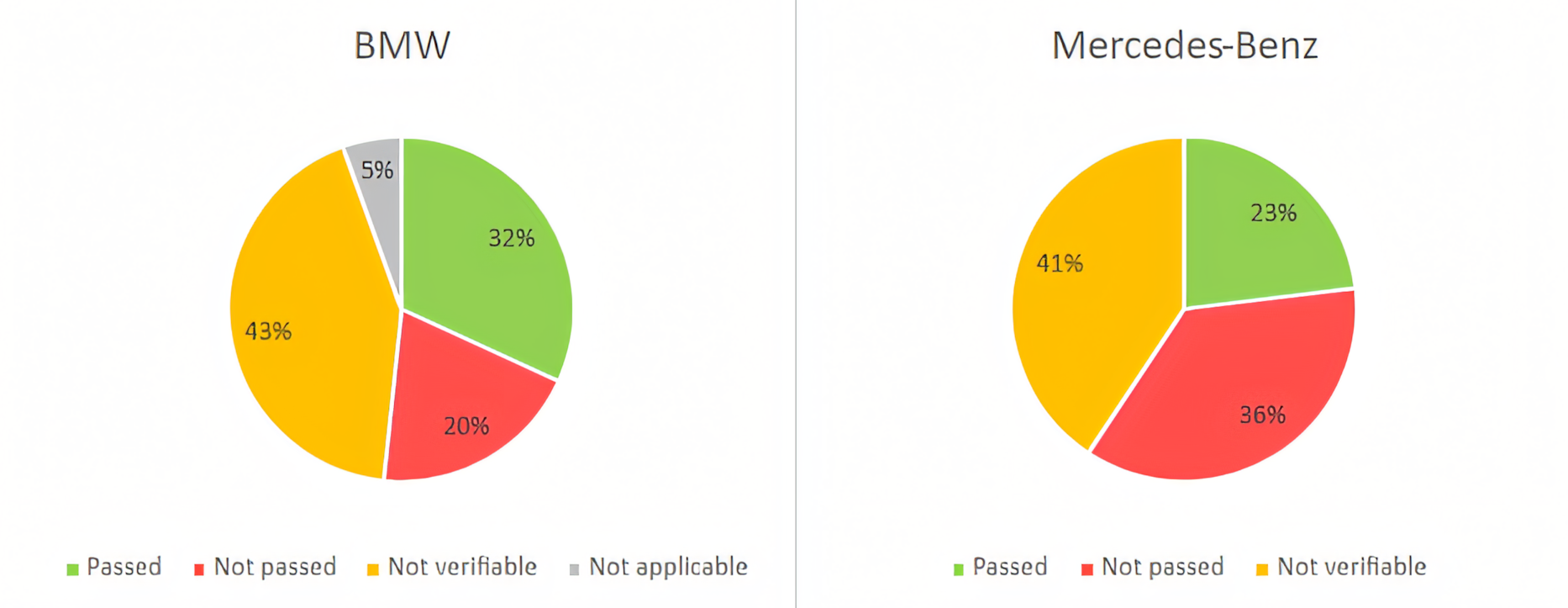}
    \caption{Pie Chart}
    \label{fig:pie-chart}
\end{figure}

Notably, BMW slightly complies with more controls than Mercedes-Benz. While this number alone seems to hint at a similar level of compliance (i.e., passed controls) between the two manufacturers, with 32\% compliance by BMW and 23\% compliance by Mercedes-Benz, we can see that this is not the case when we compare the level of non-compliance (i.e. non-passed controls). 
Mercedes-Benz's ones outweigh the ones exposed by BMW: this is partly related to the fact that Mercedes-Benz's policy is not written for the end-users, however, it does not give any instruction to group subsidiaries on how to handle certain controls. 
This does not necessarily mean that the subsidiaries are not compliant with the same controls: it just means that the data protection policy does not cover them with an appropriate level of detail. 
The high level of discrepancy between the number of controls not satisfied by BMW and those not satisfied by Mercedes-Benz can also be attributed to the fact that BMW has slightly more non-verifiable controls and a few more non-applicable controls when compared to Mercedes-Benz: adding up all these differences, we can comprehend why they have a similar level of compliance but such a diverse level of non-compliance. 
Figure \ref{fig:pie-chart} depicts the discussed information individually, for each manufacturer.

To get more details on what requirements are not satisfied, we analyse car manufacturers' compliance per topic, that is, per each chapter of the GDPR where articles from 5 to 49 are included. 
More specifically, we divided our 91 controls into four categories, for which we forged new names for brevity:

\begin{itemize}
    \item Chapter 2, Principles of Data Processing;
    \item Chapter 3, Rights of the Data Subject;
    \item Chapter 4, Responsibilities and Provisions;
    \item Chapter 5, ExtraEU Transfers of Data.
\end{itemize}

It must be observed that the other chapters of GDPR do not contain requirements for data controllers and processors, hence, we did not include them in our activity; for the same reason, we excluded Articles 11, 23, 40, 41, 42, 43, 45, 47 and 48 even if they are within the aforementioned chapters.
Finally, although Articles 31, 38 and 39 do contain requirements for data controllers and processors, assessing the manufacturers' compliance with their content would require access to internal policies, as these articles regulate the cooperation of the company with the authorities and the support that the company must guarantee to the Data Protection Officer (DPO). 
Because of this, we did not include matching controls in our checklist for these articles, hence, they do not appear among the results shown in the upcoming figures and tables.

\begin{figure}[ht]
\centering
\begin{subfigure}{.45\textwidth}
  \centering
  \includegraphics[width=.7\linewidth]{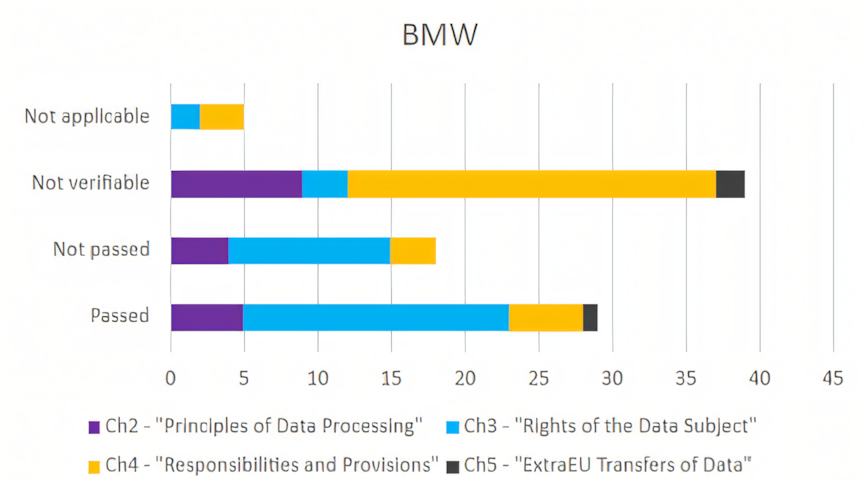}
\end{subfigure}%

\begin{subfigure}{.45\textwidth}
  \centering
  \includegraphics[width=.7\linewidth]{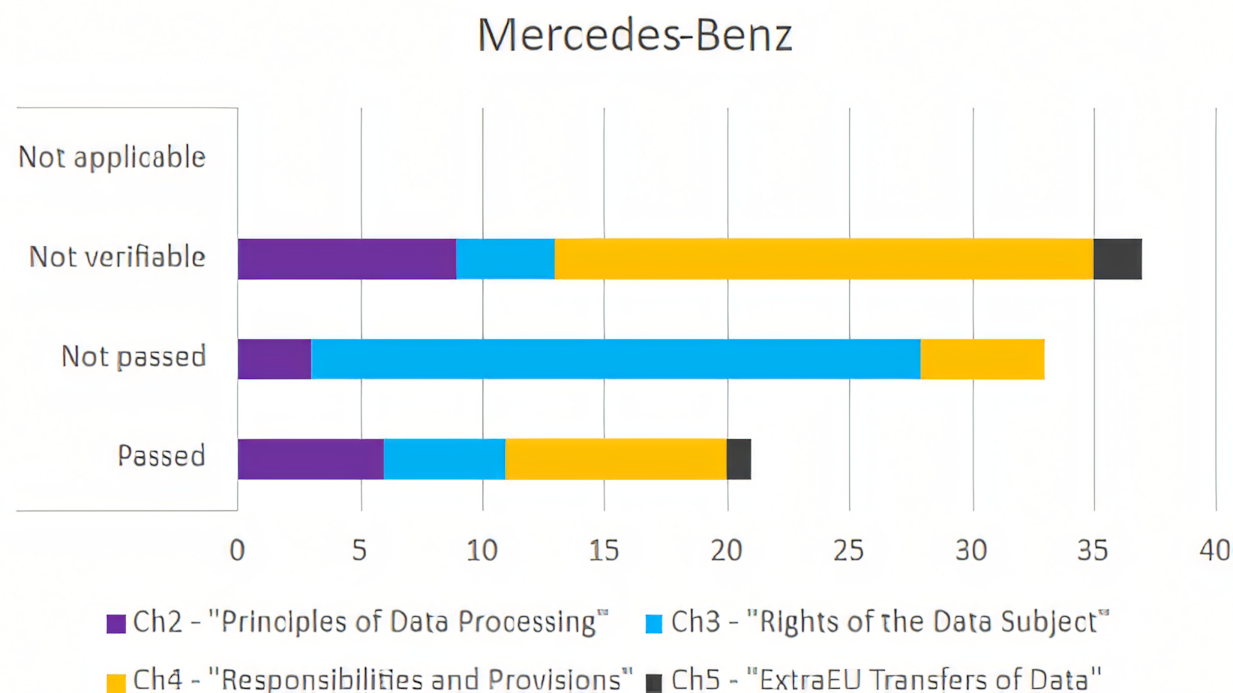}
\end{subfigure}
\caption{Gap Analysis Histograms}
\label{fig:histogram}
\end{figure}

Figure \ref{fig:histogram} shows the compliance of each manufacturer, grouped by the aforementioned chapters. Regarding ``Principles of data Processing'', we can see that BMW is compliant to 5 controls, while Mercedes-Benz is compliant to 6 of them (out of 18). 
Common gaps can be found in Articles 5, 8 and 9. Regarding Article 5, while both companies certainly make their best effort to ensure fairness of data processing, they do not explain in their privacy policies how they ensure fairness of data processing and if they have policies to periodically review the matter. 
More broadly speaking, the keyword ``fairness'' is never mentioned in either privacy policy. 
Regarding Article 8, we could not find any evidence, on the examined privacy policies, that the companies make reasonable efforts to verify the age of their users.
Finally, regarding Article 9, we could not find any additional measure implemented for ensuring the security of the processing of special categories of personal data.

For what concerns the chapter ``Rights of the Data Subject'', BMW is compliant with more than half of our controls (18 out of 32), while Mercedes-Benz is only compliant with five of them. This does not mean that Mercedes-Benz does not give its users the possibility to exercise their data rights. As usual, it only means that the examined policy does not give any instruction (to users or other companies, for their individual implementation of the related procedure) on how to properly exercise these rights. 
Common areas of non-compliance for the two companies lie in Articles 13 and 14. This is because the origin of the processed data is not always clear, so it is not always possible to understand if certain data was given by the user or was obtained by other means, which are not cited in the policies. 
Furthermore, where types of processed data are explained, their description is not always detailed (e.g., the list of processed data ends with ``etc.''), making the list incomplete.

In the chapter ``Responsibilities and Provisions'', most controls are simply not verifiable from our side, following the mere examination of the given privacy policy. This is mainly because this chapter prescribes the adoption of internal policies that we cannot verify: for example, the implementation of procedures to demonstrate compliance of the data processing to GDPR, policies for reviewing said compliance, business continuity and disaster recovery plans that are adequate to the detected level of risk, notifications of data breaches to the supervisory authorities, and much more that is not publicly available unless the privacy policy explicitly cites it. Hence, we were able to assess only a small part of the controls in this chapter: companies share non-compliance to controls in Article 25, as we could not verify that data is stored for a limited time necessary for processing in all cases.
Finally, for the chapter ``ExtraEU Transfers of Data'', only one control is verifiable from the examined privacy policies, and the two companies exhibit an identical level of compliance: more specifically, they transfer data to countries or organisations outside the EU, both where the Commission already expressed a favourable opinion on the transfer and where the Commission did not decide yet. Policies state that appropriate measures to safeguard the latter transfers are implemented, however, we could not verify why these transfers take place, e.g., if at least one of the conditions stated by Article 49 is satisfied.

\section{Limitations and Discussion}
\label{sec:disc}

This Section interprets the findings derived from the gap analysis, examines their implications, and positions them within the broader context of automotive privacy compliance.

We should remark that the gap analysis scores discussed in this paper are strictly related to the specific privacy policies under assessment. As such, the results may vary depending on the privacy policy being analysed, potential updates over time, and differences in interpretation criteria.
Therefore, it is reasonable to admit a small margin of error for each score, to account for potential discrepancies in the evaluation process.
The analysis was limited to publicly available privacy policies and did not involve a technical audit of internal company practices or systems. Therefore, some controls, particularly those requiring internal data-handling procedures, could not be verified. Furthermore, this analysis reflects the state of the privacy policies as of the time of review, and it may not fully capture future updates or changes in the legal interpretation of GDPR or other privacy regulations. Consequently, this analysis should only be taken as a contribution towards a comprehensive assessment of the overall GDPR compliance of these companies.

Within the above scope and limitations, the results of the gap analysis revealed significant non-compliance across both BMW and Mercedes-Benz, despite BMW demonstrating a slightly higher level of GDPR adherence. Table~\ref{tab:compliance-perc} summarises the compliance results in percentage form.

\begin{table}[ht]
\centering
\caption{Gap Analysis (\%)}
  \begin{tabular}{|l|c|c|}
    \hline
    & \textbf{BMW (\%)} & \textbf{Mercedes-Benz (\%)} \\
    \hline
    \textbf{Passed} & 32 & 23 \\ \hline
    \textbf{Not Passed} & 20 & 36 \\ \hline
    \textbf{Not Verifiable} & 43 & 41 \\ \hline
    \textbf{Not Applicable} & 5 & 0 \\ \hline
\end{tabular}
\label{tab:compliance-perc}
\end{table}

In detail, BMW passed 32\% of the GDPR controls, while Mercedes-Benz passed only 23\%. This suggests that BMW is somewhat more proactive in addressing GDPR requirements, though neither brand meets even half of the relevant controls. The gap of 9\% between the two manufacturers may indicate that BMW’s privacy policy is marginally more robust, as indicated in their slightly better adherence to GDPR standards.
Mercedes-Benz shows a significantly higher rate of non-compliance, with 36\% of the controls not being met, compared to 20\% for BMW. This nearly twofold difference highlights that Mercedes-Benz’s privacy policy falls short in more critical areas of GDPR compliance.
Both car manufacturers have a high percentage of controls that are classified as “Not Verifiable” (43\% for BMW and 41\% for Mercedes-Benz). This suggests that a large portion of the privacy policies lack sufficient detail or transparency to assess full compliance. The similarity in these percentages may also indicate a broader industry issue, where automotive companies may be deliberately vague or lack precise language regarding key GDPR controls.
BMW had 5\% of controls deemed “Not Applicable”, while none of the controls for Mercedes-Benz fell into this category. The presence of non-applicable controls for BMW may suggest a broader privacy policy scope, which could include areas less relevant to the automotive context. However, this also reflects an area where the analysis might vary depending on interpretations of GDPR applicability.

Furthermore, the analysis showed that both manufacturers struggle with GDPR compliance, particularly in areas related to user transparency, data origin disclosure, and user rights enforcement. For instance, both companies failed to fully comply with Article 13, which requires clear communication of the origin of personal data. In many cases, the policies used vague language, such as listing “data categories” followed by “etc.”, without fully specifying the types of data being collected or how the data was obtained.
This lack of transparency directly impacts user control over personal data, raising concerns about how well users can exercise their rights under GDPR, such as the right to erasure or access. These gaps suggest that drivers are not made fully aware of the scope of the data being collected and of how to request deletion of their data, thus making it harder to mitigate risks associated with privacy issues, e.g., unauthorised data access, which could materialise during cyber warfare especially.

In addition, the findings also highlight broader implications in the context of cybersecurity. Given the extensive data collection and connectivity features of modern vehicles, insufficient GDPR compliance may increase vulnerabilities during cyber warfare. In fact, in scenarios where drivers’ data (e.g., location or vehicle operation data) are exploited, gaps in data protection could turn connected vehicles into tools for targeted attacks, compromising personal safety.
In particular, the failure to implement specific controls related to the right to erasure or restrictions on data transfers outside the EU (Article 49) may expose drivers to heightened risks if their data is used in cross-border operations. As we could not verify the compliance with Article 49 in both the car brands, it is noteworthy to stress the importance of clearer communication of GDPR controls by automotive manufacturers. 
In fact, this lack of transparency exacerbates the risks, as it becomes difficult to assess how well users’ personal data is truly protected in the event of cyberwarfare.

Finally, it can be concluded that the results of this gap analysis align with previous findings in the literature, as Section~\ref{sec:related-work} showed, yet with the inclusion of the dimension of compliance with specific GDPR requirements. 
In fact, while other works focused on general cybersecurity and privacy threats aboard smart cars, this work frames those risks in terms of legal and regulatory compliance, ultimately offering a more holistic assessment of both technical and legal shortcomings, particularly relevant for the implications that may derive in the context of cyber warfare.

The results also suggest that the automotive industry as a whole may be lagging in terms of privacy compliance. 
A possible lesson learned could be that automotive manufacturers need to adopt more rigorous internal policies, such as regular audits of privacy practices and better alignment between legal and technical teams, to address the existing gaps.

\section{Conclusions}
\label{sec:conc}

This paper advanced a gap analysis of two leading automotive manufacturers, BMW and Mercedes-Benz, to assess the extent to which their privacy policies align with the stringent requirements of GDPR. By examining compliance at the requirement level, it became clear that both brands exhibit significant areas of non-compliance, with Mercedes-Benz lagging slightly behind BMW. The primary issues identified include non-verifiable compliance, incomplete data origin disclosure, and limited instruction on the exercise of user rights.

The findings suggest that both manufacturers face considerable challenges in ensuring GDPR compliance, particularly in terms of transparency and user data control. Given the increasing data collection by connected vehicles, these gaps could expose drivers to privacy risks and potential cyberattacks, which may become even more crucial in the context of cyberwarfare. 
In fact, the failure to protect user data adequately may allow modern vehicles to be weaponised against their users in the event of a cyber conflict. This highlights the need for more robust privacy policies and clearer communication of data practices, in support of stronger enforcement mechanisms to protect users.

Furthermore, the automotive industry's current state of privacy compliance appears to be inadequate.
Regular audits, better integration between legal and technical teams, and ongoing updates to privacy policies can be essential steps toward closing the gap and ensuring compliance with data protection laws.

Our future work looks at expanding this study with gap analyses on the other top 10 car manufacturers in 2023 (and potentially 2024), to identify wider areas of non-compliance. In addition, automating the analysis of privacy policies could provide more efficient and scalable methods for ongoing compliance assessments. As the automotive industry increasingly integrates autonomous driving technologies and AI, understanding and mitigating the risks associated with data misuse in cyberwarfare contexts will be crucial to protect drivers and maintain trust in connected vehicles.

\section*{Acknowledgment}

This work acknowledges financial support from: PRIN 2022 MUR project E53D23008220006-FuSeCar.

\bibliographystyle{alpha}
\bibliography{references}

\end{document}